\documentclass{aa}
\input epsf
\def\spose#1{\hbox to 0pt{#1\hss}}
\def\mdot{\spose{\raise 7.0pt\hbox{\hskip 5.0pt{\char '056}}}M}

\begin{document}
\thesaurus{08.23.2; 08:06.3; 08.09.2 (Sk41); 11.13.1; 11.19.3} 

\title{Wind properties of Wolf-Rayet stars at low metallicity: Sk\,41 (SMC)
\thanks{Based on observations collected 
at the European Southern Observatory, La Silla, Chile. Proposal
Nos. 61.D--0680 and 63.H--0683.}}
\author{Paul A. Crowther}

\offprints{P.A. Crowther (pac@star.ucl.ac.uk)}
\institute{Department of Physics and Astronomy, University College London,
           Gower Street, London WC1E 6BT, U.K.}
\date{received 4 November 1999; accepted ???}
\authorrunning{Paul A. Crowther}
\titlerunning{WR stars in the SMC}
\maketitle
\begin{abstract}
The stellar properties of Sk\,41 (AB4, WN5h), the only known
single Wolf-Rayet star in the SMC, are derived from
ultraviolet (IUE), optical (AAT) and near-IR (NTT) spectroscopy.
Contrary to expectations, the stellar properties of 
Sk\,41 are typical of equivalent WN stars in the Galaxy 
and LMC, with $T_{\ast}$$\sim$42kK, log($L/L_{\odot}$)=5.7,
$v_{\infty}$=1300 km\,s$^{-1}$, 
$\dot{M}/\sqrt{f}$=3$\times$10$^{-5} M_{\odot}$yr$^{-1}$,
and H/He$\sim$2 by number, where $f$ is the volume filling
factor. The stellar luminosity of Sk\,41 is 50\% below the 
minimum value predicted by single star evolutionary models 
at the metallicity of the SMC.

Emission line luminosities of He\,{\sc ii} $\lambda$4686 and
C\,{\sc iv} $\lambda\lambda$5801-12 in SMC WR stars are not 
systematically lower than their Galactic and LMC counterparts. 
From 43 late-type and 59 early-type WN stars, 
log~$L_{\lambda}^{\rm HeII}$= 36.0 erg\,s$^{-1}$ 
and 35.8 erg\,s$^{-1}$, respectively, while
log~$L_{\lambda}^{\rm CIV}$=36.5 erg\,s$^{-1}$.
from 25 early-type WC stars. This new calibration has application 
in deriving WR populations in young starburst galaxies.

Synthetic WN models are calculated with identical parameters
except that metal abundances are varied. Following the Smith 
et al. WN classification scheme, CNO equilibrium models reveal 
that earlier spectral types are predicted at lower metallicity, i.e. 
WN3--4 at 0.04$Z_{\odot}$ versus WN6 at 1.0$Z_{\odot}$. 
This provides an explanation for the 
trend towards earlier WN spectral types at low metallicity.


\end{abstract}

\section{Introduction}

Galaxies containing the youngest starbursts, with ages of a few Myr, are 
known as Wolf-Rayet galaxies (Vacca \& Conti 1992) since they
show the spectroscopic signature of large numbers of WR stars.
At present, in excess of 140 WR galaxies have been discovered
(Schaerer et al. 1999), including examples at very 
low metallicity, $Z$, such as I\,Zw18 with $Z$=0.02$Z_{\odot}$ (De 
Mello et al. 1998), probably typical of star forming galaxies 
in the early universe.

\begin{table}
\caption[]{Revised catalogue of SMC WR stars, updated from 
Azzopardi \& Breysacher (AB, 1979) to include MDV1 (Morgan et al. 1991) 
and exclude the O3If/WN6 star AB2 (AzV39a, Walborn 1977).
Spectral types follow Smith et al. (1996) for WN stars or Crowther 
et al. (1998) for the WO star. A 
foreground extinction of $E_{\rm B-V}$=0.03 mag 
is adopted, together with the SMC extinction law of Bouchet et al. (1985).}
\label{table1}
\begin{flushleft}
\begin{tabular}{l@{\hspace{0mm}}c@{\hspace{2mm}}r@{\hspace{1mm}}c
@{\hspace{2mm}}l@{\hspace{1mm}}r@{\hspace{2mm}}l@{\hspace{2mm}}r
@{\hspace{2mm}}r}
\hline\noalign{\smallskip}
Name & AB & Sk & AzV & Sp Type & $m_v$& $E_{\rm B-V}$ &$M_{v}$& $M_{v}$ \\ 
     &    &    &     &         & sys  & mag           & sys   & WR \\ 
\hline\noalign{\smallskip}
     & 1  &    & 2a & WN3+O4   &15.2 &0.26 &-4.6&-3.6 \\% & WN3+O4   O= -4
     & 3  &    & 60a& WN4+O4   &14.7 &0.17 &-4.8&-4.1 \\% & WN3+O4   O= -4
     & 4  &41  & 81 & WN5h     &13.4 &0.11 &-5.8&-5.8 \\% & WN6-A    
MVD1 &    &    &    & WN3+O5:  &15.6 &0.10?&-3.7&-3.7 \\% & WN2.5+O5: 
HD5980$^{\ast}$
     & 5  & 78 & 229& WN3--4+O7I  &11.6 &0.07 &-7.5&-6.5 \\% &  WN3= -6.5 O7I=-7.0
     &    &    &    & WN11h    & 8.6 &0.07 &-10.5&-10.5 \\% & LBV=-10.6
     &    &    &    & WN6(h)   &10.4 &0.07 &-8.7&-8.7 \\% & WNL= -6.6
     & 6  &108 & 332& WN3+O6.5I&12.1 &0.07 &-6.8&-5.2 \\% & WN3+O6.5I O= -6.5
     & 7  &    &336a& WN3+O7   &13.1 &0.11 &-6.2&-5.2 \\% & WN3+O7   O=-5.6
Sand 1& 8 & 188&    & WO3+O4V  &12.7 &0.05 &-6.1&-5.0 \\% & WO4+O4V  O=-5.6
\noalign{\smallskip}\hline
\end{tabular}
\end{flushleft}
$\ast$: Three entries are given for HD\,5980, correponding to circa 1981
(pre-outburst, Torres-Dodgen \& Massey 1988), September 1994 (mid-outburst, 
Heydari-Malayeri et al. 1997), and December 1997 (post-outburst, own dataset).
\end{table}

The O star content of starburst galaxies is derived from 
nebular Balmer line fluxes, while WR populations are obtained 
from comparing broad He\,{\sc ii} $\lambda$4686 and 
C\,{\sc iv} $\lambda$5801-12 emission line fluxes with calibrations
of individual Galactic and/or LMC WR stars
(Schaerer \& Vacca 1998).

Radiative driven wind theory predicts that 
$\dot{M} \propto Z^{0.5}$ (Kudritzki et al. 1989). 
Incorporating this effect into massive star 
evolutionary models implies that the minimum initial mass 
star reaching the WR phase through single star evolution 
is predicted to be a function of $Z$; 
25$M_{\odot}$ in the Galaxy, 35$M_{\odot}$ in the LMC and 45$M_{\odot}$
in the SMC (Maeder 1997). Although WR winds are considered to 
be driven by radiation pressure, their mass-loss rates are assumed to be 
independent of metal content in current evolutionary calculations.

Do WR stars have comparable mass-loss properties and emission line
strengths within various metallicity regions? Attempts to establish
differences between the properties of O and WR 
stars in the Galaxy and LMC have proved inconclusive 
(e.g. Puls et al. 1996; Crowther \& Smith 1997), since the metal 
content of the LMC differs from the Solar neighbourhood only by 
a factor of $\sim$2.

Consequently, the SMC represents our closest neighbour in 
which to establish whether the stellar 
properties of O and WR stars are affected by low metal
content (12+log~O/H=8.1; Russell \& Dopita 1990). Walborn 
et al. (1995), Puls et al. (1996) and Prinja \& Crowther (1998) 
demonstrated that SMC O-type stars indeed reveal lower mass-loss 
rates and slower winds, except amongst very early O giants 
and supergiants. Comparisons for 
Wolf-Rayet stars are more problematic, since they are very rare in 
the SMC, and almost entirely binaries. Omitting O3\,If/WN stars, 
the LMC contains 125 bona-fide WR stars (Breysacher et al. 1999), 
in contrast to only  eight WR stars in the SMC, which are 
listed in Table~\ref{table1}. Crowther (1999) discusses the 
influence of metallicity dependent
WR mass-loss rates on stellar spectra and ionizing flux distributions.

HD\,5980 has recently received considerable interest 
because of a Luminous Blue Variable (LBV)-type eruption in one of its 
component 
stars (Barba et al. 1995; Koenigsberger et al. 1998; Moffat et al. 1998), 
although its variability and multiplicity hinders the reliability
of spectroscopic analysis. 

In order to avoid uncertainties caused by binarity,
a study of the only known single WR star in the SMC, Sk\,41 (AB4), as
discussed by Moffat (1988), is presented here, using the non-LTE 
line blanketed model atmosphere code of Hillier \& Miller (1998). 
Its  stellar properties are compared with counterparts in the Galaxy and 
LMC, and the role of metal content on spectral appearance is investigated. 
In addition, the line luminosities of He\,{\sc ii} 
$\lambda$4686 and C\,{\sc iv} 
$\lambda\lambda$5801-12 in SMC WR stars are compared with 
Galactic and LMC WR stars, and a new calibration is derived.

\begin{figure*}
\epsfxsize=12.0cm \epsfbox[20 200 520 780]{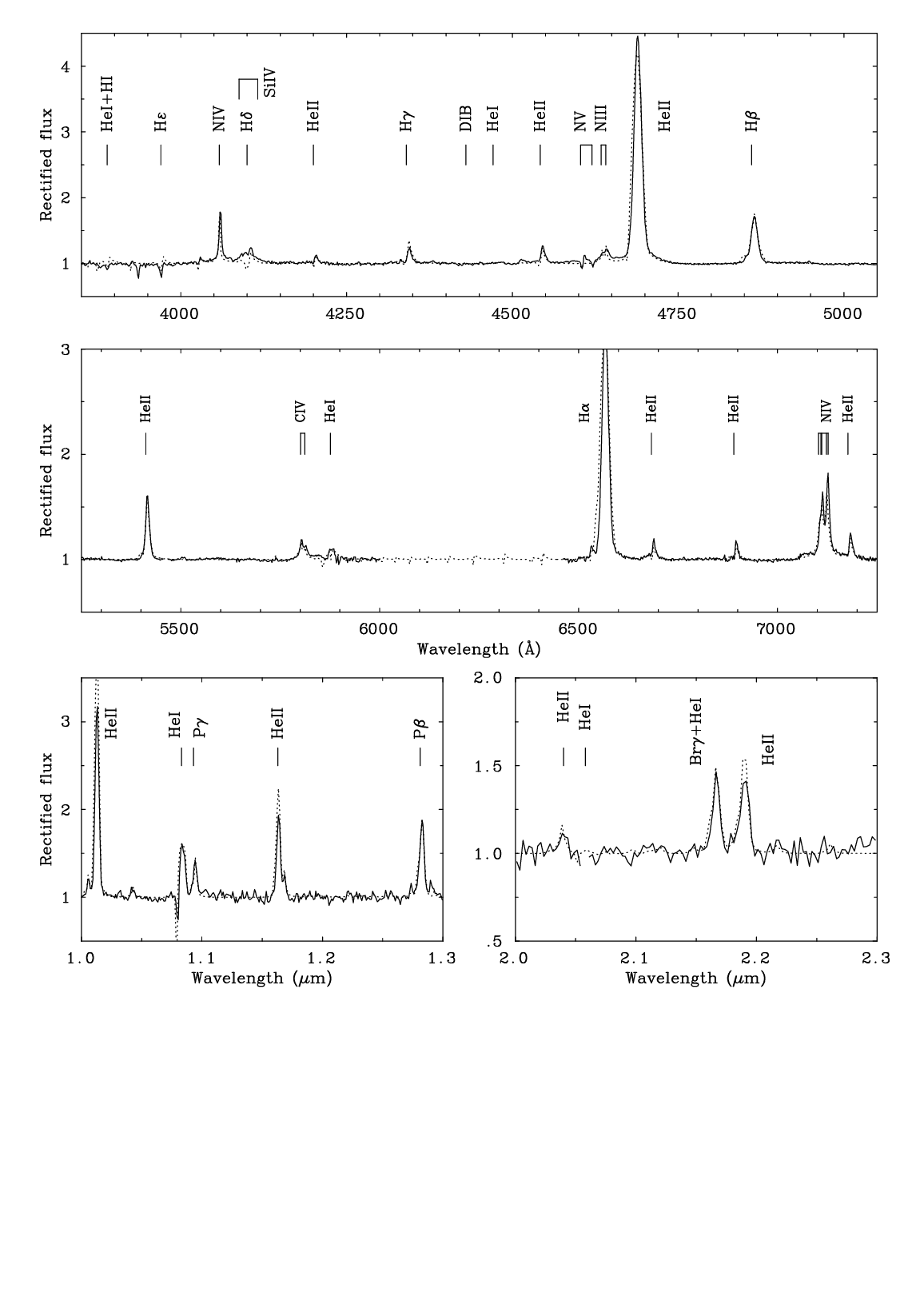}
\caption{Comparison between rectified optical (AAT/RGO) and
near-IR (NTT/SOFI) spectroscopy of Sk\,41 (solid) and synthetic 
spectra (dotted).}
\label{fig1_rect}
\end{figure*}

\section{Observations}

UV, optical and near-IR spectroscopy of Sk\,41 has been obtained from 
the International Ultraviolet Exporer (IUE) archive, the
3.9m Anglo-Australian Telescope (AAT) and the European Southern Observatory 
3.5m New Technology Telescope (NTT), respectively. Following 
Smith et al. (1996),
a spectral classification of WN5h is obtained for Sk~41, in comparison
to previous classifications of WN6-A (Walborn 1986) or WN4.5 
(Conti et al. 1989). Although it could be claimed that a WN5h+abs
classification is more appropriate from the Smith et al. definition,
we adopt WN5h because of the close similarity of Sk\,41 with other 
WN5h stars, such as HD\,65865 (WR10) and R136a1 (BAT99-108).

\subsection{Optical spectroscopy}

An optical spectrogram of Sk\,41 was obtained at the AAT
during 1992 November 3--6, using the RGO spectrograph, 25cm camera, 
Tektronix CCD (1024 $\times$ 1024, 24$\mu$m pixels), 1200V and 1200B 
gratings, plus a slit width of 2$''$. The measured spectral resolution 
in the extracted spectra is 1.6--1.8\AA, using the FWHMs of the Cu-Ar 
arc spectra. Four settings covered 3670--6005\AA, plus 6455--7263\AA.
A standard reduction was carried out as discussed by Crowther \& 
Smith (1997). Subsequent analysis made use of {\sc dipso} 
(Howarth et al. 1998). In addition, a low resolution flux 
calibrated spectrum of Sk\,41 from Torres-Dodgen \& Massey (1988) 
has been used.

\subsection{Near-IR spectroscopy}

Long slit, near-IR spectroscopy of Sk\,41 was acquired with
the NTT, using the Son OF Isaac (SOFI) instrument, a 1024x1024 pixel
NICMOS detector, and low resolution IJ (GRB) and HK (GRR)
gratings on 1 Sept 1999. The spectral coverage was 0.94--1.65$\mu$m and 
1.50--2.54$\mu$m, respectively, with dispersions of 7.0\AA/pix and
10.2\AA/pix. The 0.6 arcsec slit provided a 2 pixel spectral resolution 
of 14--20\AA. The total integration time was 960 sec at each grating 
setting. Atmospheric calibration was achieved by observing HD~10747 (B3V) 
immediately before or after Sk\,41, at an close airmass (within 0.03). 
Similar observations of HD\,2002 (F5V) permitted a relative flux 
correction, using a $T$=6,500K model atmosphere normalized to V=8.13 mag.

A standard extraction and wavelength calibration was carried out with
{\sc iraf}, while {\sc figaro} (Shortridge et al. 1999) and {\sc dipso} were 
used for the  atmospheric and flux calibration, first artificially removing 
stellar hydrogen features from the B3V spectrum. Convolving our fluxed
spectra with suitable filters indicates J$\sim$13.6, H$\sim$13.5, and K$\sim$13.3 
mag.

\subsection{UV spectroscopy}

A short wavelength (SWP48325), high resolution (HIRES) {\it IUE} 
observation of Sk~41 was retrieved from the World Data Centre at 
the Rutherford Appleton Laboratory. This was obtained 
on 7 August 1993,
with an exposure time of 20,340 sec and was reduced using {\sc iuedr} (Giddings
et al. 1996).
In addition, three final archive (IUEFA), low resolution, large aperture 
datasets were obtained from STScI. Two short wavelength (SWP6195,
SWP107270) and one long wavelength (LWR5359) spectra were obtained
between 15 Aug 1979 and 2 Dec 1980, with exposure times of 900, 1200
and 960 sec, respectively.

\section{Spectroscopic analysis}

The model calculations are based on the iterative technique 
of Hillier (1987, 1990) which solves the transfer equation in 
the co-moving frame subject to statistical and radiative equilibrium, assuming
an expanding, spherically-symmetric, homogeneous and static atmosphere. 
Allowance is made for line blanketing and clumping following the
formulation of Hillier \& Miller (1998). 

Calculations consider detailed model atoms of 
H\,{\sc i}, He\,{\sc i-ii}, C\,{\sc iv}, N\,{\sc
iii-v}, O\,{\sc iii-vi}, Si\,{\sc iv} and 
Fe\,{\sc iv-vii} (see Dessart et al. (2000) for details
of the source of atomic data). Weak transitions of iron 
($gf \le 10^{-4}$) have been excluded without affecting
the emergent spectrum. In total, 1027 full levels 
and 7680 non-LTE transitions are simultaneously 
considered. These are combined into 275 `super levels', with 
solely the populations of the super level calculated in the 
rate equations. Populations of individual atomic levels 
are then determined by assuming that  it has 
the same departure coefficient as the super level to which 
it belongs.

In line with 
recent iron determinations for O stars in the SMC (Haser et al. 1998),
abundances of elements other than hydrogen and helium are fixed at
0.2 $\times$ Solar (Si=0.02\% by mass, Fe=0.03\% by mass) or 0.2 $\times$
Solar CNO-equilibrium values (C=0.008\%, N=0.3\%, O=0.005\% by mass).

\begin{figure}
\epsfxsize=8.8cm \epsfbox[50 400 500 750]{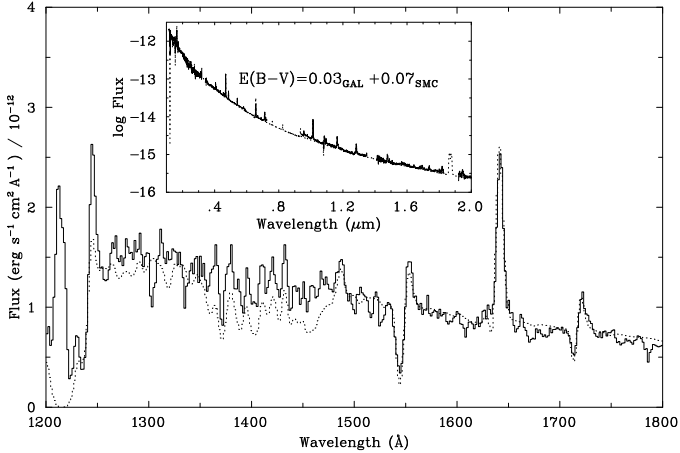}
\caption{Comparison between UV (IUE/LORES) spectrophotometry 
of Sk\,41 (solid), dereddened by $E_{\rm B-V}$=0.03$_{\rm Gal}+0.07_{\rm SMC}$,
with synthetic spectra degraded to the IUE resolution (dotted), allowing
for interstellar Ly$\alpha$ absorbtion. 
The inset box compares the model energy distribution with UV-optical-IR observations.}
\label{fig1}
\end{figure}

The analysis technique follows that of Crowther et al. (1995), such 
that diagnostic
lines of He\,{\sc i} ($\lambda$10830), He\,{\sc ii} ($\lambda$5412)
and H\,{\sc i} (H$\beta$+He\,{\sc ii} $\lambda$4859) are chosen to derive
the stellar temperature, mass-loss rate, luminosity and hydrogen content.
The mass-loss rate is actually derived as the ratio $\dot{M}/\sqrt{f}$,
where $f$ is the volume filling factor. This can be constrained by
fits to the electron scattering wings of the helium line profiles.
A wind velocity of $\sim$1300~km\,s$^{-1}$ is obtained from IUE/HIRES C\,{\sc
iv} $\lambda$1548-51 observations, although the S/N of this dataset is low.
A standard $\beta$=1 velocity law is adopted, in 
order to provide consistency with recent analyses. However, 
recent evidence  indicates that early WR winds accelerate more slowly 
(e.g. Lepine \& Moffat 1999), so we have also investigated the effect of
a slow $\beta \sim$10 law on derived stellar parameters. For Sk\,41,
a slow velocity law reveals a mass-loss rate that is systematically lower 
by $\sim$25\%, with the stellar temperature and luminosity barely affected.

\subsection{Stellar properties of Sk\,41}

Rectified optical (AAT/RGO) and near-IR (NTT/SOFI) spectroscopy 
of Sk\,41 is compared with our synthetic spectra in 
Fig.~\ref{fig1_rect}. Overall, agreement is excellent for most
H\,{\sc i}, He\,{\sc i-ii}, N\,{\sc iii-iv} and C\,{\sc iv} transitions.
The fit to the He\,{\sc ii} $\lambda$4686 electron scattering wing is excellent
using a clumped model with a volume filling factor of $f$=0.1. 
Note that N\,{\sc v} $\lambda$4603--20 emission is not predicted, while 
the H$\delta$ region, containing Si\,{\sc iv} $\lambda\lambda$4088-4116 and 
N\,{\sc iii} $\lambda\lambda$4097--4103 is underestimated.
Consistency between optical and near-IR fits in Sk\,41 is good,
although near-IR He\,{\sc ii} transitions, such as 1.01$\mu$m and 2.19$\mu$m,
are $\sim$25\% too strong, based on fits to optical
He\,{\sc ii} transitions. Similar results were obtained for strong-lined
WNE stars by Crowther \& Smith (1996). 

Fig.~\ref{fig1} compares the dereddened energy
distribution of Sk\,41 with model predictions, using
a distance of 60.2\,kpc to the SMC (Westerlund 1997). The 
Seaton (1979) extinction law is adopted for the foreground 
extinction, assumed to be $E_{B-V}$=0.03 mag. In addition, 
the Bouchet et al. (1985) SMC law as parameterised by Calzetti 
(priv. comm.) is used for internal SMC extinction. 
$E_{B-V}=0.07$ mag is required to match the synthetic model 
(Fig.~\ref{fig1}, inset box). SMC interstellar Lyman$\alpha$ 
absorbtion has been accounted for, assuming
$n$(H\,{\sc i})=10$^{21.8}$ cm$^{-2}$ (Fitzpatrick 1985).

Synthetic filter photometry from 
Torres-Dodgen \& Massey (1988) indicates
$b$=13.22 mag for Sk\,41, so the absolute b-magnitude of 
Sk\,41 is $M_{b}=-$6.1 mag. Agreement between the IUE dataset 
and the synthetic spectra (degraded to the IUE/LORES resolution) is 
reasonable, although imperfect in the $\lambda\lambda$1250--1500\AA\ region, 
dominated by transitions of Fe\,{\sc v}.

The derived stellar parameters for the WN5h star 
are $T_{\ast}$=42kK, $R_{\ast}$=13.6$R_{\odot}$, log($L/L_{\odot}$)=5.7, 
H/He=2.25 by number, and 
$\dot{M}/\sqrt{f}$=2.8$\times$10$^{-5}$$M_{\odot}$yr$^{-1}$.
The bolometric correction of this model is $-$3.9 mag, while the
predicted number of ionizing photons shortward of H\,{\sc i} 
$\lambda$911 and He\,{\sc i} $\lambda$504 are 
log $Q_{0}$=49.45 s$^{-1}$ and log $Q_{1}$=48.77 s$^{-1}$, 
respectively.

\subsection{Comparison with WN stars in the Galaxy and LMC}

\begin{figure}
\epsfxsize=8.0cm \epsfbox[0 0 500 775]{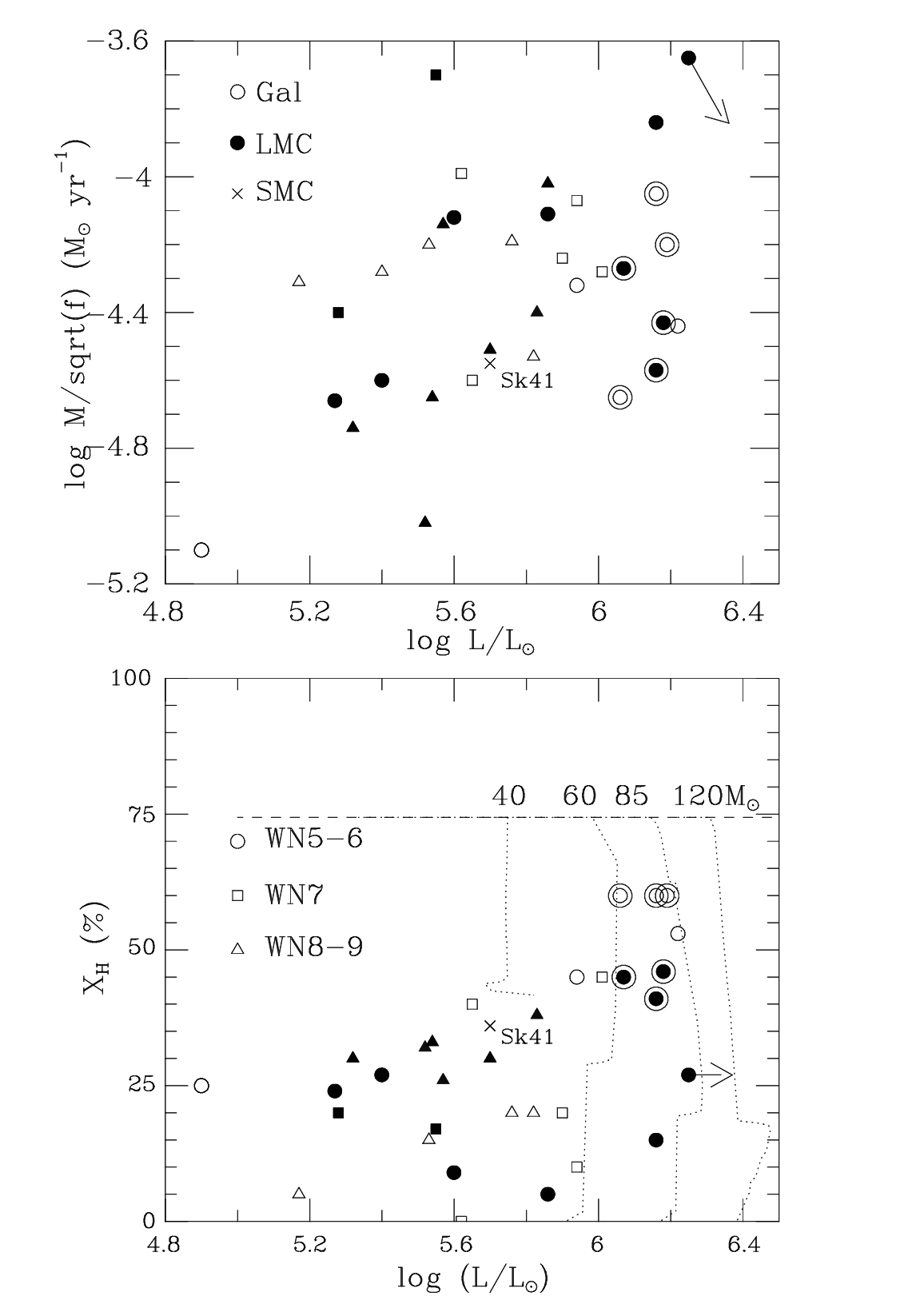}
\caption{Comparison between derived stellar properties of Sk\,41 (WN5h,
cross) with WNL stars at known distance in the Galaxy 
(open symbols) and LMC (filled symbols). Stellar parameters 
follow helium diagnostics, except for six members of giant
H\,{\sc ii} regions, 
for which nitrogen diagnostics were adopted
(concentric circles, Crowther \& Dessart 1998). An 
arrow indicates the effect of using nitrogen instead of 
helium diagnostics for HD\,38282 (BAT99-118, WN6h), although
no shift is appropriate for Sk\,41. Evolutionary predictions 
at 0.2$Z_{\odot}$ are shown as
dotted lines (Meynet et al. 1994), with the main 
sequence (appropriate for the SMC) indicated by a 
dot-dashed line.}
\label{fig2}
\end{figure}

Historically, WN5--6 spectral types have been considered solely
as early WN (WNE) stars. In contrast, we assign a WNL 
status for those WN5--6 stars which contain atmospheric hydrogen 
following Smith et al. (1996), for consistency with evolutionary 
definitions. Consequently, the stellar properties of Sk~41 are shown in 
the upper panel of Fig.~\ref{fig2} (cross), together with a large 
sample of Galactic and LMC WN5--9 stars.
Stellar parameters of the other stars are taken from Hamann et al. 
(1993, 1995), Crowther et al. (1995), Crowther \& Smith (1997) and 
Crowther \& Dessart (1998). Previous approaches were identical except that 
line blanketing and clumping were neglected, and nitrogen stellar 
diagnostics were  employed by Crowther \& Dessart (1998).
The effect of selecting nitrogen, rather than 
helium diagnostics is indicated as an arrow for HD\,38282 
(BAT99-118, WN6h) in Fig.~\ref{fig2}, although 
from the previous discussion no shift is appropriate for Sk\,41.

The mass-loss rate of Sk~41 compares closely to other WNL stars
with similar luminosities. A homogeneous model implies a 
wind  performance number, $\mdot v_{\infty}/(L_{\ast}/c)$, of 
3.6 for Sk\,41, which compares closely with the three luminous 
WN5h stars in R136a (Crowther \& Dessart 1998). Considering a volume 
filling factor of $f$=0.1 reduces the wind  performance number for 
Sk\,41 to approximately unity. Although the wind velocity of 
Sk\,41 is a factor of two lower than that of the R136a stars, it is 
comparable to the sole Galactic WN5h star HD\,65865 (WR10), for 
which $v_{\infty}$=1500 km\,s$^{-1}$ (Hamann et al. 1993).

It remains to be successfully demonstrated whether WR winds in
high metallicity environments can be driven solely by multiple 
scattering (Schmutz 1997). Therefore, the wind properties
of Sk\,41 provide an excellent challenge at low metallicities.

\begin{figure}
\epsfxsize=8.8cm \epsfbox[20 300 470 780]{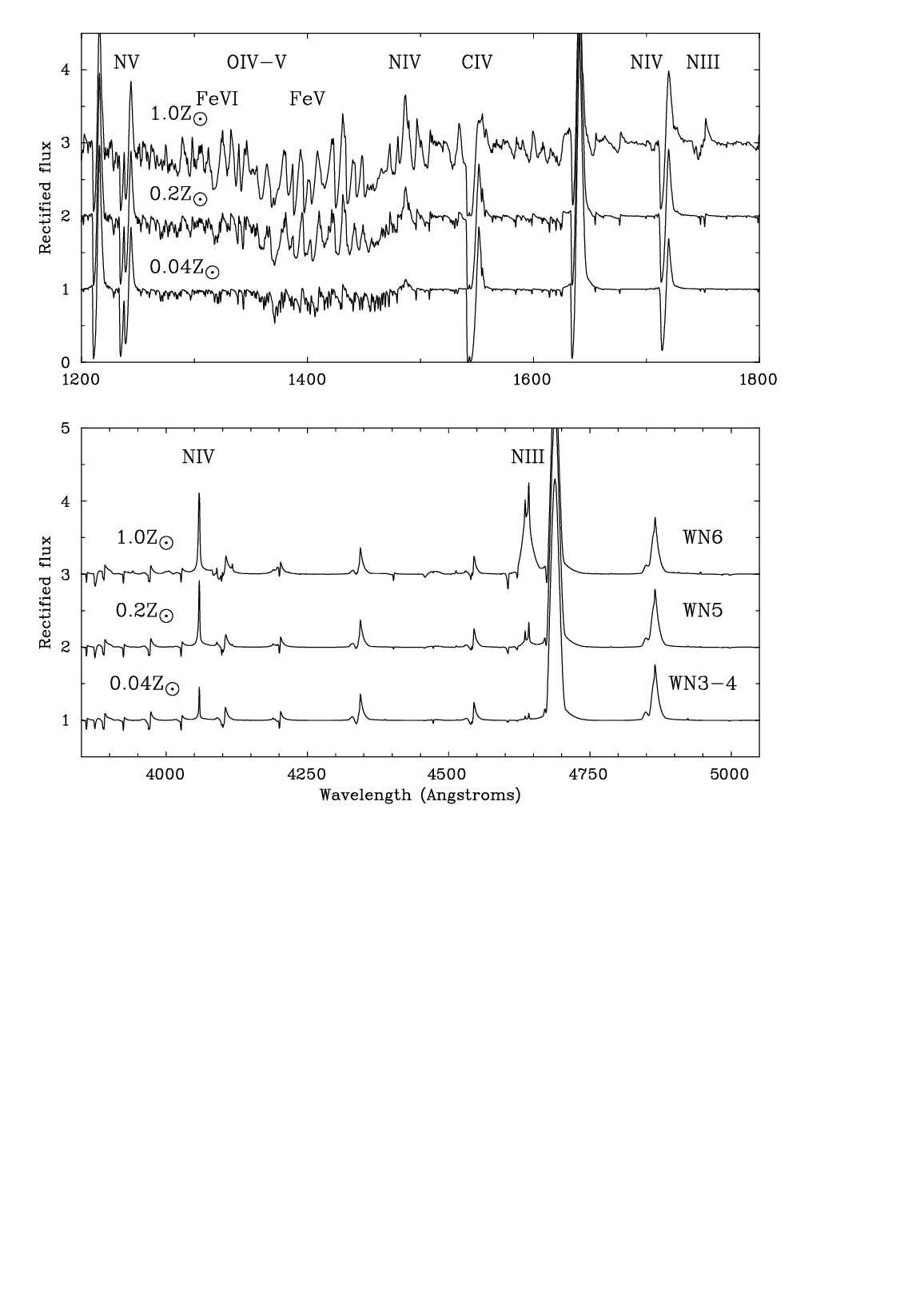}
\caption{Synthetic UV and optical spectra of WN models 
for which the metal content is varied between solar (1.0$Z_{\odot}$),
SMC (0.2$Z_{\odot}$) and 0.04$Z_{\odot}$.
In all cases assumed stellar parameters are $T_{\ast}$=42\,kK, 
log($L/L_{\odot}$)=5.7, $v_{\infty}$=1300 km\,s$^{-1}$, 
$\dot{M}$=9$\times$10$^{-6} M_{\odot}$yr$^{-1}$, and $f$=0.1}
\label{fig1_metal}
\end{figure}

\subsection{Evolutionary model predictions at low metallicity}

Since Sk\,41 is apparently single, how do its properties compare with
theoretical expectations for single massive stars at SMC metallicities
(Meynet et al. 1994)? In the lower panel of Fig.~\ref{fig2} we show the
luminosity and hydrogen content (in mass fraction) for a large sample of
Galactic and LMC WNL stars, as well as Sk~41 (again indicated by cross). 
The latest evolutionary tracks for metallicities appropriate
to the SMC (0.2$Z_{\odot}$, Meynet et al. 1994) are superimposed
with initial masses of 40, 60, 85 and 120$M_{\odot}$. The minimum 
initial mass that is predicted to
progress through to the WN phase has 45--50$M_{\odot}$, with 
log($L/L_{\odot})\sim$5.9, and a WN lifetime of $\le$100,000 yr.
Therefore, Sk~41 has a stellar luminosity which is 50\% lower 
than the minimum that is predicted by evolutionary models. 
Since it is unlikely that Sk~41 is a disrupted binary (its
radial velocity is typical of SMC stars), our results appear
to identify a deficiency in present evolutionary models at low
metallicity. Either less massive stars can advance to the WR stage,
or the stellar luminosity decreases as the star reaches advanced 
evolutionary phases. These are completely determined by the 
mixing and the previous evolutionary phases, which remain poorly known.

\section{Effect of metal content on spectral types of  WN stars}

If the mass-loss properties of WN stars in different environments
are unaffected by metallicity, do differences in 
metal content have any effect on the emergent spectral appearance? 
To investigate this, we have fixed the stellar properties of Sk\,41, and 
calculated additional models in which the metal content is varied by 
a factor of five, to solar (1.0$Z_{\odot}$) or 0.04$Z_{\odot}$,
equivalent to twice the metal content of I\,Zw18 which is known to 
host WR stars (e.g. De Mello  et al. 1998). 

CNO elements are fixed at equilibrium WN values, appropriate for each 
environment (Meynet et al. 1994), such that the nitrogen 
content varies from 0.06\% to 1.5\% by mass. The atmospheric structures
of these models are identical, except that since their outer 
wind temperatures are dependent on the cooling of 
the wind through metal resonance lines (Hillier 1988).
At a radius of 10$R_{\ast}$, equivalent to $\tau_{\rm Ross}\sim$0.01,
the wind temperature is $T^{\rm wind}_{e}$=23kK, 18.5kK and 16kK, for 
0.04$Z_{\odot}$, 0.2$Z_{\odot}$ and 1.0$Z_{\odot}$, respectively.

Fig.~\ref{fig1_metal} shows  synthetic UV and optical WN spectra for 
each case. In the UV, differences between the models are dominated
by the strength of Fe\,{\sc v-vi} features and the appearance
of N\,{\sc iii} $\lambda\lambda$1748--52 in the Solar metallicity model.
Differences in blanketing play a minor role in the Lyman ionizing 
flux distribution of these models, although shortward of 
the O$^{+}$ edge (353\AA) the higher blanketing of the 1.0$Z_{\odot}$ 
model predicts a factor of three times fewer ionizing photons than 
the 0.04$Z_{\odot}$ case.

{}From Fig.~\ref{fig1_metal}, the optical region reveals that 
solely low excitation lines, such as N\,{\sc iii} $\lambda$4634--41 and 
He\,{\sc i} $\lambda$4471, are enhanced at high metallicities.
He\,{\sc ii} $\lambda$4686 is essentially unaffected, as is N\,{\sc iv} $\lambda$4058 
due to its complex line formation mechanism, despite the factor
of 25 decrease in nitrogen content. Differences for N\,{\sc iii} 
$\lambda$4634--41 are principally due to abundance effects. Lines 
formed in the outer wind, such as He\,{\sc i}, are  sensitive to 
wind cooling, such that the emission equivalent width of 
He\,{\sc i} $\lambda$10830 decreases from 59\AA\ at 1.0$Z_{\odot}$, 
to 35\AA\ at 0.2$Z_{\odot}$ and 26\AA\ at 0.04$Z_{\odot}$.

\begin{table*}
\caption[]{Comparison between mean line luminosities (in units of 10$^{35}$ erg\,s$^{-1}$)
and mean FWHM (\AA) of He\,{\sc ii} $\lambda$4686 in Galactic, LMC and SMC WN stars,
and C\,{\sc iv} $\lambda\lambda$5801--12 in WC and WO stars. Note that we 
consider WN2--4 plus hydrogen-free WN5--6 stars as WNE stars, and WNL otherwise.}
\label{table3}
\begin{flushleft}
\begin{tabular}{l@{\hspace{7mm}}
r@{\hspace{2mm}}r@{\hspace{1mm}}c@{\hspace{1mm}}r@{\hspace{7mm}}
r@{\hspace{2mm}}r@{\hspace{1mm}}c@{\hspace{1mm}}r@{\hspace{7mm}}
r@{\hspace{2mm}}r@{\hspace{1mm}}c@{\hspace{1mm}}r@{\hspace{7mm}}
r@{\hspace{2mm}}r@{\hspace{1mm}}c@{\hspace{1mm}}r}
\hline\noalign{\smallskip}
He\,{\sc ii} $\lambda$4686/& \multicolumn{4}{c}{Galaxy} & 
\multicolumn{4}{c}{LMC} & \multicolumn{4}{c}{SMC} & \multicolumn{4}{c}{Total} \\
C\,{\sc iv} $\lambda$5801 & $L_{\lambda}$  & $\sigma$ & FWHM & N & $L_{\lambda}$  & $\sigma$ & FWHM & N & 
$L_{\lambda}$  & $\sigma$ & FWHM & N & $L_{\lambda}$  & $\sigma$ & FWHM & N \\
\hline\noalign{\smallskip}
WNE$\dag$     & 7.6  & 5.0& 33&18 & 5.8 & 5.6& 28&  37 & 2.7 & 0.8 &25& 4       & 6.1& 5.4 &29& 59\\
WNL           & 7.6  & 6.5& 17&17 &13.5 &14.2& 16&  25 & 7.8 & --  &13& 1       &11.0&11.9 &16& 43\\
\hline\noalign{\smallskip}
WCE           & 17.0 & 10.5 &47 & 9 &40.3  & 18.1& 64 & 16 & --   &  -- &    & 0 & 31.9 & 19.4 & 58& 25\\
WO$^\ast$     &  2.6 &  --  &150 & 1 &15.8  &  --&103 & 1  & 32.3 &  -- &92 & 1 & 14.9 & 11.1 &104& 4\\
\noalign{\smallskip}\hline
\end{tabular}
\end{flushleft}
$\dag$: HD\,5980 has been excluded (see Sect.~\ref{line_lumin});
$\ast$: The WO total includes DR1 in IC1613 (Kingsburgh \& Barlow 1995).
\end{table*}

Therefore, although {\it identical} physical parameters are adopted in each 
WN model, {\it earlier} spectral types are obtained at {\it lower} 
metallicity. Following the scheme of Smith et al. (1996), whose
spectral classification diagnostics were specifically chosen to 
avoid metallicity effects, the spectral
type resulting from our sample of models ranges from WN6 
at 1.0$Z_{\odot}$ to WN3--4 at 0.04$Z_{\odot}$ (the latter
depending on the selection of diagnostic lines).
This effect may contribute to the trend towards early WN types at 
low metallicities. Recall that 100\% of SMC WN stars have spectral 
types of WN2--5, in contrast to 78\% in the LMC and 46\% in the Galaxy.

\section{Line strengths of Wolf-Rayet stars}

Unfortunately, definitive results are not possible from our comparison 
between the physical parameters of Sk\,41 with the wide range of parameters
observed in Galactic and LMC counterparts. Conti et al. (1989) have compared the 
emission equivalent widths of SMC WR stars with those of the Galaxy
and LMC, which revealed that their emission lines 
are relatively weak. Massey et al. (1987) and Armandroff \& Massey (1991)
extended these comparisons to other Local Group galaxies.
However, their conclusions are affected by binarity, or line-of-sight 
companions. 
Instead, absolute line luminosities are presented here, since they
are unaffected by a binary companion, and also used to determine 
the WR  stellar content of starburst galaxies. Comparisons with absolute 
visual magnitudes provide a superior indication of wind emission strengths.
Conti \& Massey (1989) provide a comparison between line fluxes
and absolute visual magnitudes for LMC and Galactic WN stars, while
Massey \& Johnson (1998) present a similar comparison for LMC, 
SMC and M33 WN stars.

\subsection{Sample of Wolf-Rayet stars}

He\,{\sc ii} $\lambda$4686 (for WN2--9) and 
C\,{\sc iv} $\lambda$5801-12 (for WC4--6 and WO) line luminosities have
been measured in a large sample of Galactic, LMC and SMC WR systems, 
restricting the former to stars of known distance, generally through 
cluster/association membership. These two lines were selected since they 
are amongst the strongest features in all spectral types, 
and are commonly used to assess WR populations
in starburst regions (Schaerer \& Vacca 1998). 

The majority of our 
measurements were taken from the 
Torres-Dodgen \& Massey (1988) atlas, except where superior resolution 
fluxed datasets are available to us, obtained with either AAT/RGO,
MSO 74inch/coude or MSO 2.3m/DBS during Dec 1991--Dec 1997 
(e.g. Crowther \& Smith 1997). 

Absolute visual magnitudes have been calculated as follows.
For WN stars, interstellar reddenings were generally obtained from
weighted averages of Schmutz  \& Vacca (1991), Morris et al. (1993), Hamann
et al. (1993), Hamann \& Koesterke (1998) and our own determinations
based on comparing dereddened optical and UV (IUE) datasets with theoretical 
WN flux distributions. Interstellar extinction laws follow
Seaton (1979) for the Galaxy, Howarth (1983) for the LMC, and 
Bouchet et al. (1985) for the SMC. WC interstellar reddenings 
were weighted averages of Smith et al. (1990ab), Morris et al. (1993), 
Koesterke \& Hamann (1995), 
Kingsburgh et al. (1995), Gr\"{a}fener et al. (1998), and again our own 
determinations. Individual measurements are available upon request to
the author. Spectral types are taken from Smith et al. (1990b,
1996) and Crowther et al. (1998). Magellanic Cloud
distances are taken from Westerlund (1997), namely 51.2kpc 
and 60.2kpc for the LMC and SMC, respectively.

Absolute magnitudes of WR stars in binaries have been corrected for the 
presence of their OB companion. Where possible, previously estimated 
light ratios are adopted (e.g. Smith et al. 1996). Otherwise, standard OB 
calibrations are used. In some cases, such as AzV~2a (WN3+O4), O-type
calibrations (Conti et al. 1983; Crowther \& Dessart 1998) exceed 
the systemic absolute magnitude, so either the spectral type of 
their companions is in error, or they are sub-luminous. 
For these systems, we generally adopt $M_{\rm V}$(O)=$-$4.0 mag,
with the exception of MDV1, for which no correction was applied (Table~\ref{table1}).

\begin{figure}
\epsfxsize=8.8cm \epsfbox[50 275 500 775]{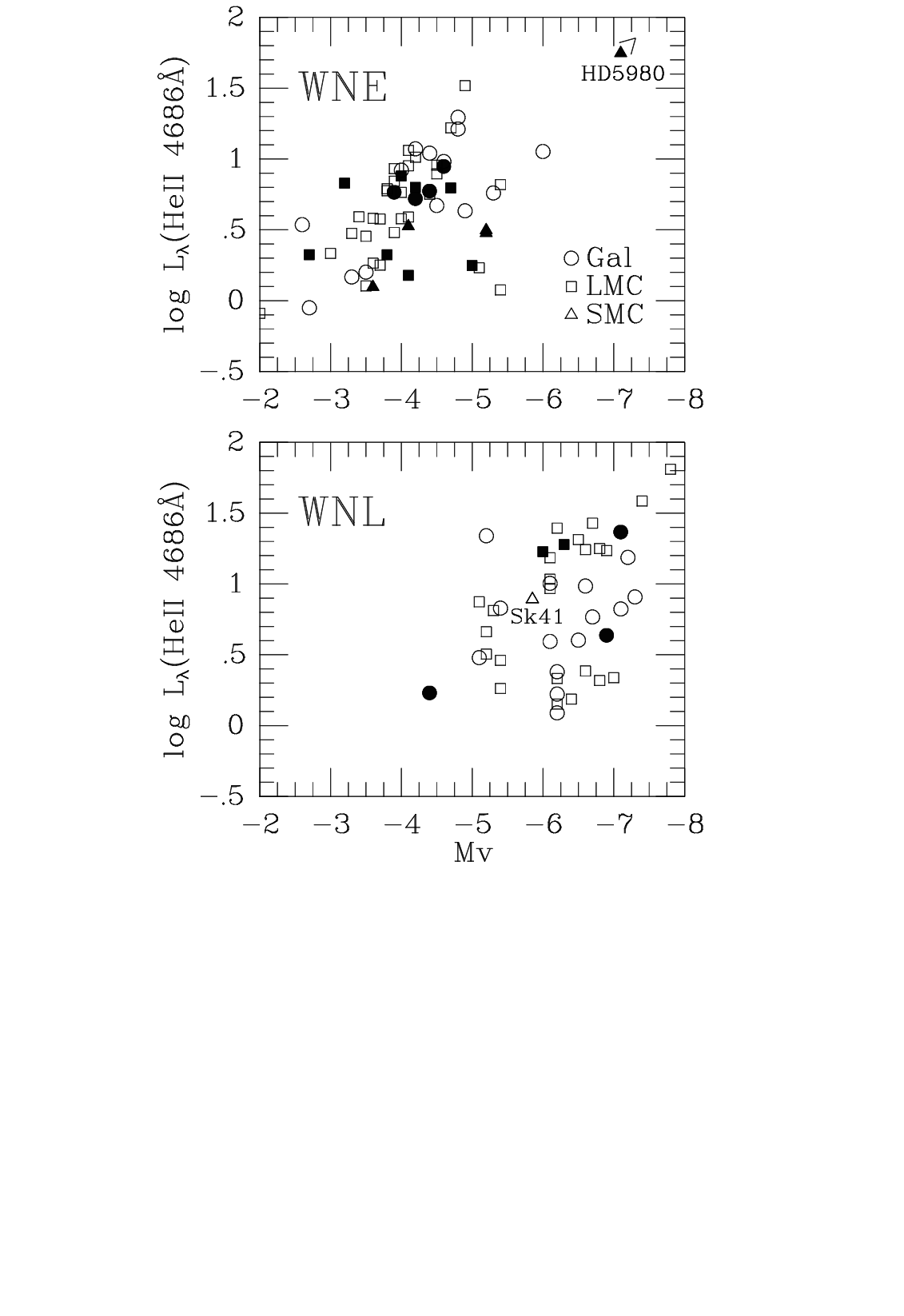}
\caption{Line luminosities (10$^{35}$ erg\,s$^{-1}$)
of He\,{\sc ii} $\lambda$4686 in WN stars. Galactic stars
are represented by circles, LMC stars by squares and SMC stars 
by triangles. Single stars are open symbols, with binaries 
indicated by filled in symbols.}
\label{fig3}
\end{figure}

\begin{figure}
\epsfxsize=8.8cm \epsfbox[50 525 500 775]{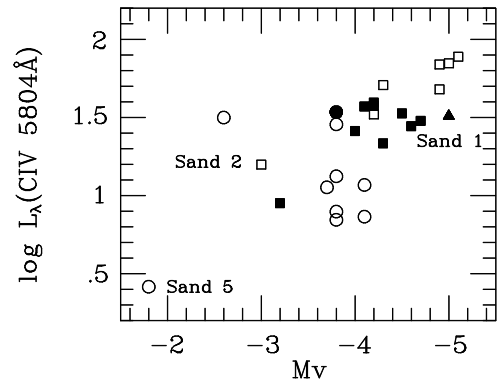}
\caption{Line luminosities (10$^{35}$ erg\,s$^{-1}$)
of C\,{\sc iv} $\lambda$5801-12 in WCE (WC4--6) and WO stars. 
Galactic stars are represented by circles, LMC stars by 
squares and SMC stars  by triangles. Single stars are 
open symbols, with filled-in symbols for binaries.
The identity of WO stars is shown.}
\label{fig4}
\end{figure}

\subsection{Line luminosities}\label{line_lumin}

In Figs.~\ref{fig3}--\ref{fig4} we present line luminosities of Galactic, 
LMC and SMC WN and WC stars. Although the scatter is 
large, we confirm the trend towards higher 
line luminosities for WN stars with increasing $M_{v}$, previously identified
by Conti \& Massey (1989) for the LMC, and find a similar correlation
for WCE stars. Absolute magnitudes of single stars (open symbols) 
are more reliable than members of binaries (filled-in symbols).

{}Table~\ref{table3} presents mean line luminosities for Galactic,
LMC and SMC WN and WC stars, and the combined sample. Note that the 
formal standard deviations are poor indicators, given the small 
numbers involved and  non-gaussian distribution. From the Table, 
the line luminosities of WN stars in the SMC are 
apparently a factor of 2--3 times lower than their LMC and Galactic 
counterparts. However, line luminosities of SMC WNE stars do  not
do not fall systematically below Galactic or LMC WNE stars with
comparable (low) absolute magnitudes as shown in Fig.~\ref{fig3}. 
Similarly, Sk\,41 has a luminosity that is within $\sim$20\% of mean 
Galactic and LMC values, if the remarkable WN5--6 stars in 
LMC and Galactic giant H\,{\sc ii} regions are excluded 
(e.g. Crowther \& Dessart 1998).

HD\,5980 is omitted from the SMC sample shown in Table~\ref{table3}, 
since its line luminosity is from multiple components, highly variable and 
indeed of uncertain origin\footnote{Moffat et al. (1998) attribute 
pre-outburst line emission to  material formed in the shocked 
region between two early type components, rather than a single WNE star}. 
The He\,{\sc ii} $\lambda$4686 luminosity 
of HD\,5980 was 5.59$\times$10$^{36}$ erg\,s$^{-1}$ in 1981, prior 
to outburst, when its spectral type was WN3--4. Observations obtained 
in Dec 1994, shortly after outburst, when the spectral type was 
WN8, revealed an even more  remarkable He\,{\sc ii} line 
luminosity of 2.26$\times$10$^{37}$ erg\,s$^{-1}$, which is a
factor of four greater than any other WN system from our sample,
albeit with an equally impressive absolute visual magnitude
(Table~\ref{table1}).

Amongst WC/WO stars, comparisons between the SMC and other 
galaxies are hindered by the absence of WC stars in the SMC. Amongst the 
WO stars with known distances, Sand~5 (WR142, WO2) in 
the Galaxy, Sand~2 (BAT99-123, WO3) in the LMC, 
and Sand~1 (AB8, WO3+O) in the SMC, it is Sand~1 which has 
the largest C\,{\sc iv} $\lambda$5801 luminosity. However, 
attempts to draw 
conclusions are severely hindered by the very low numbers involved. 
To illustrate this, DR1 (WO3) in the SMC-like IC\,1613 has a 
C\,{\sc iv} line luminosity which is a factor of 
three times lower than Sand~1 (Kingsburgh \& Barlow 1995).

Therefore, WO and WN stars in the SMC do not have systematically lower 
line luminosities than their higher metallicity counterparts. Therefore,
mean values of our entire sample may be determined, and are
indicated in Table~\ref{table3}. How do mean line luminosities compare 
with previous determinations? Smith et al. (1990a)  
obtained 3.3$\times$10$^{36}$ erg s$^{-1}$ 
from observations of 5 LMC WC4 stars, which is confirmed here, 
based on a much larger sample of 25 Galactic and LMC WCE stars. 
More recently, Schaerer \& Vacca (1998) obtained 
(5.2$\pm$2.7)$\times$10$^{35}$ erg s$^{-1}$ from 26 WNE stars and 
(1.6$\pm$1.5)$\times$10$^{36}$ erg s$^{-1}$ from 19 WNL stars. From
59 WNE stars, a 20\% higher mean results here, while 43 WNL stars
indicate a 30\% lower calibration for late WN stars. 

Note that application of these mean luminosities will lead 
to an overestimate of the true WN population in starburst regions by 
a factor of $\sim$2--3 if luminous WN5--6h stars dominate the WR 
signature, as is the case in 30 Dor and NGC\,3603. Conversely, if 
the properties of constituent WNE stars are more typical of SMC stars,
where close binary evolution probably plays an important role, the 
actual population would be underestimated by a similar factor. 

\subsection{Line widths of Galactic, LMC and SMC WR stars}

If WR stars in the SMC have slower winds than their Galactic and 
LMC counterparts
due to lower metallicity, we expect that their line widths will also be
systematically lower.

However, in contrast to line fluxes, FWHM are sensitive to 
observational resolution. For example, AB7 has FWHM(He\,{\sc ii} $\lambda$ 4686)=32\AA\
from low resolution spectroscopy of Torres-Dodgen \& Massey (1988), 
in contrast to 26\AA\ from medium resolution AAT/RGO data. This observational
limitation also complicates the Smith et al. classification scheme, which
assigns `b' for broad lined stars if FWHM(He\,{\sc ii} $\lambda 4686)\ge$30\AA\ 
(see also Conti 1999). Consequently, we have measured 
FWHM(He\,{\sc ii}) for a WR sample obtained with a
uniform medium resolution of $\sim$2\AA). Where these are unavailable, corrections 
to measurements from Torres-Dodgen \& Massey (1988) datasets are made, using
stars in common to both datasets as calibrators.

From Table~\ref{table3}, we find that the mean FWHM(He\,{\sc ii} 
$\lambda$4686) for WNE stars does decrease at lower metallicity,
from 33\AA\ in the Galaxy to 25\AA\ in the SMC. However, since the SMC  WR
population is so low, taking into consideration the dependence 
of FWHM on individual spectral types, no unambiguous conclusions 
may be drawn. For example, LMC WCE stars reveal systematically higher 
FWHM (C\,{\sc iv} $\lambda$5801--12) than Galactic stars since 
the LMC WC population is solely represented by WC4 stars, which are rare
in our Galaxy. The situation is further complicated in our Galaxy
since Schild et al. (1990) and Armandroff \& Massey (1991) 
identified a correlation between 
FWHM(C\,{\sc iv} $\lambda$5801--12) and galacto-centric distance 
for WCE stars in our Galaxy and M33, attributed to a metallicity gradient.
Amongst WO stars, we find an apparent trend towards narrower lines at 
lower metallicity, again based on a very small sample.
             
\section{Conclusions}

Contrary to expectations, it appears that both the 
stellar properties and line luminosities of Wolf-Rayet stars in 
the SMC do not differ significantly from their counterparts in 
higher metallicity galaxies. Therefore, the reliability of studies of
WR starburst galaxies at low metallicity based on template 
Galactic or LMC stars is supported here. However, individual 
spectral types may differ by up to a factor of $\sim$5 from 
mean WN, WC or WO line luminosity calibrations. Therefore,
the question of what Wolf-Rayet flavours are produced in 
different star forming regions becomes relevant.

One added complication is that close binary evolution probably 
plays a major role in the formation of SMC WR stars, which may affect 
their physical properties. This is probably not the case for 
regions undergoing powerful bursts of massive star formation. 
It would be very useful to compare stellar properties and line 
luminosities of individual WR stars in low metallicity starbursts 
with the present calibrations, for which IC10 represents an 
ideal candidate (Massey \& Armandroff 1995).

\begin{acknowledgements}
Thanks to John Hillier for providing his current stellar atmospheric
code, and to Werner Schmutz for permitting the use of new SOFI 
datasets in this work. I appreciate Daniela Calzetti providing a 
parameterization of the Bouchet et al. SMC law, and Peter Conti, 
Nolan Walborn and Orsola De Marco for a careful reading of this 
manuscript. PAC is funded by a University Research Fellowship
of the Royal Society. The support of the staff from the 
Anglo-Australian, European Southern and Mount Stromlo Observatories 
is greatly appreciated. This work has made use of the 
SIMBAD database, operated at the CDS, Strasbourg, France. 
\end{acknowledgements}

\end{document}